\documentclass
[superscriptaddress,secnumarabic,amssymb,amsmath,nobibnotes,aps,prd,showkeys,showpacs,nofootinbib,onecolumn,12pt]{revtex4}%
\usepackage{graphicx}
\usepackage{subfigure}
\usepackage{epsf}
\usepackage{bm}
\usepackage{amsmath}
\usepackage{amsfonts}
\usepackage{amssymb}%
\providecommand{\U}[1]{\protect\rule{.1in}{.1in}}

\newcommand{\be}{\begin{equation}}
\newcommand{\ee}{\end{equation}}

\newcommand{\mincir}{\raise
-3.truept\hbox{\rlap{\hbox{$\sim$}}\raise4.truept\hbox{$<$}\ }}
\newcommand{\magcir}{\raise
-3.truept\hbox{\rlap{\hbox{$\sim$}}\raise4.truept\hbox{$>$}\ }}

\usepackage{hyperref}
\hypersetup{
    colorlinks=true,
    linkcolor=red,
    citecolor=blue,
    filecolor=magenta,      
    urlcolor=cyan,
    pdftitle={New analytic solutions in f(R)-Cosmology from Painlev\'{e} analysis},
    pdfpagemode=FullScreen} 

\begin{document}
\title{New analytic solutions in $f\left(  R\right)  $-Cosmology from Painlev\'{e} analysis}
\author{Genly Leon}
\email{genly.leon@ucn.cl}
\affiliation{Departamento de Matem\'{a}ticas, Universidad Cat\'{o}lica del Norte, Avda.
\ Angamos 0610, Casilla 1280 Antofagasta, Chile }
\affiliation{Institute of Systems Science, Durban University of Technology, PO Box 1334,
Durban 4000, South Africa}
\author{A. Paliathanasis}
\email{anpaliat@phys.uoa.gr}
\affiliation{Institute of Systems Science, Durban University of Technology, PO Box 1334,
Durban 4000, South Africa}
\affiliation{Instituto de Ciencias F\'{\i}sicas y Matem\'{a}ticas, Universidad Austral de
Chile, Valdivia 5090000, Chile}
\author{P.G.L. Leach}
\email{leachp@ukzn.ac.za}
\affiliation{Institute of Systems Science, Durban University of Technology, PO Box 1334,
Durban 4000, South Africa}
\begin{abstract}
Using the singularity analysis, we investigate the integrability properties and existence of analytic solutions in  $f\left(  R\right)$-cosmology. Specifically, for some power-law $f\left(  R\right)  $-theories of particular interest, we apply the ARS algorithm to prove if the field equations possess the Painlev\'{e} property. Constraints for the free parameters of the power-law models are derived, and new analytic solutions are derived, expressed in terms of Laurent expansions.

\end{abstract}
\date{\today}
\maketitle

\section{Introduction}

\label{sec1}

Determining exact and analytic solutions to describe
dynamical systems is essential for analysing systems in all
physics and applied mathematics areas. Nowadays, numerical techniques are
applied widely for the analysis of differential equations. However, the
derivation of exact and analytic solutions for a set of nonlinear differential
equations gives an elegant way to analyse the dynamical system.
Arscott expressed in a lovely way in the preface of his book \cite{ar1} the necessity for the search for analytic solutions and not relying on numerical
methods; Arscott wrote: \textquotedblleft\lbrack...] fall back on numerical
techniques savours somewhat of breaking a door with a hammer when one could, with a little trouble, find the key\textquotedblright.

Einstein's General Relativity provides a set of nonlinear differential
equations, known as Einstein's field equations, which relate the physical
space and the matter source. In cosmological studies, the field equations are
reduced to a system of nonlinear ordinary differential equations that still
in the case of a vacuum, can not be solved easily. Some cosmological
solutions describing vacuum geometries are presented in \cite{v2,v3,v4,v5}. In
the simplest cosmological scenario, for which the physical space is described by
the spatially flat Friedmann--Lema\^{\i}tre--Robertson--Walker (FLRW)
geometry, with or without the cosmological constant term, the nonlinear
ordinary differential equations can be linearised under the application of a
point transformation \cite{nl1}. The linearisation of this field
equations are ensured by the existence of the maximum number of point
transformations under which the field equations are invariant. The symmetry analysis is a systematic way to determine invariant functions for the dynamical system, which has been widely applied in gravitational physics and cosmology
\cite{sm1,sm2,sm3,ap1,sym4}. Furthermore, symmetry analysis is the main
approach for the investigation of the integrability properties and the
construction of analytic solutions for the field equations in modified and
extended theories of gravity, see for instance \cite{ns1,ns2,ns4,ns5,ns6} and
references cited therein.

Another systematic approach for constructing analytic solutions applied in gravitational physics is the singularity analysis or the Painlev\'{e} analysis. The singularity analysis is a powerful tool for determining the integrability of differential equations. Singularity analysis is associated with the French school led by Painlev\'{e} in the previous century \cite{pa1,pa2,pa3,pa4}. Which followed the application to the determination of the third integrable case of Euler's equations for a spinning top by Kowalevskaya \cite{Kowlevskaya89a}. For an extended mathematical discussion of
the Painlev\'{e} analysis, we refer the reader to \cite{cont1}. This procedure has been the primary mathematical tool in the debate on the integrability properties of the Mixmaster Universe (Bianchi IX), see the discussion in \cite{CotsakisLeach, Con1, Demaret,bun1}. However, it has been widely applied in other gravitational models; see, for instance,
\cite{miritzis,ftAn,fqAn,cots11,cots2, cots,sbs}.

In this work, we are interested in the application of singularity analysis
for the construction of new analytic solutions in $f\left(  R\right)
$-gravity \cite{Buda} within the cosmological scenario of a spatially flat
FLRW universe. $f\left(  R\right)  $-gravity is a fourth-order theory of
gravity in which the Einstein-Hilbert Action is modified such that the Ricci scalar $R~$ in the Action
Integral is replaced by a function $f(R)$ \cite{Sotiriou,odin1}.
The main idea behind modifying the gravitational Action Integral is
that the new degrees of freedom introduced by the modification drive the
cosmological dynamics to explain the cosmological observations
\cite{capQ}. The quadratic theory of gravity  $f\left(
R\right)  =R+qR^{2}$ \cite{qua1,qua2,qua3}, also known as the Starobinsky model of inflation
\cite{star}, is one of the most known $f\left(  R\right)  $ models. This
specific theory fits the cosmological observations and provides a theoretical mechanism to explain the inflationary parameters \cite{planck2015}. The
integrability properties of this specific theory were found recently with the
application of the singularity analysis in \cite{anst1}, while with the same
approach, the analytic solution for the power-law theory $f\left(  R\right)
=R+qR^{n}$ \ was determined in \cite{anst2}. In the following, we consider
other models of $f\left(  R\right)  $ theory, and we seek if the field
equations possess the Painlev\'{e} property and if the cosmological solution can
be written in terms of Laurent expansions. Such analysis provides 
important information about the $f\left(  R\right)  $-cosmology and the
general behaviour of the dynamics. The plan of the paper is as follows.

Section \ref{sec2} presents the field equations in the context of
$f\left(  R\right)  $-cosmology. The main properties and definitions for the
singularity analysis are given in Section \ref{sec3}. The new results of this
study are presented in Section \ref{sec4}, where we present new analytic
solutions in $f\left(  R\right)  $-theory for specific functional forms of the
$f\left(  R\right)  $ function. Finally, we draw our conclusions in Section \ref{sec5}.

\section{$f\left(  R\right)  $-Cosmology}

\label{sec2}

In $f\left(  R\right)  $-gravity the gravitational Action Integral is
\cite{Buda}
\begin{equation}
S=\int dx^{4}\sqrt{-g}\left[  \frac{1}{2k}f\left(  R\right)  +L_{m}\right],
\label{ac.01}%
\end{equation}
where $R$ is the Ricci scalar for the background geometry with metric
$g_{\mu\nu}$, $g$ is the determinant of the metric tensor, $L_{m}$ is the
Lagrangian function for the matter source, and we use units in which $k\equiv8\pi G$. Because the
Ricci scalar includes second-order derivatives for a nonlinear function $f$,
it is easy to infer that the field equations which follow from (\ref{ac.01})
are of fourth order. In addition, for a linear function $f$, that is
$f^{\prime\prime}=0$; $f^{\prime}=\frac{df}{dR}$, the Einstein-Hilbert Action
is recovered, and the gravitational theory is reduced to General Relativity.

Variation of the Action Integral (\ref{ac.01}) with respect to the metric
tensor produces the gravitational field equations, they are%
\begin{equation}
f^{\prime}R_{\mu\nu}-\frac{1}{2}fg_{\mu\nu}-\left(  \nabla_{\mu}\nabla_{\nu
}-g_{\mu\nu}\nabla_{\sigma}\nabla^{\sigma}\right)  f^{\prime}=kT_{\mu\nu},
\label{ac.02}%
\end{equation}
in which $R_{\mu\nu}$ is the Ricci tensor, and $T_{\mu\nu}=\frac{\partial
L_{m}}{\partial g^{\mu\nu}}$ is the energy-momentum tensor for the matter source.

Equations (\ref{ac.02}) can be written in the equivalent expression
\begin{equation}
R_{\mu\nu}-\frac{1}{2}Rg_{\mu\nu}=k_{\mathrm{eff}}\left(  T_{\mu\nu}+T_{\mu
\nu}^{f\left(  R\right)  }\right),  \label{ac.03}%
\end{equation}
where $k_{\mathrm{eff}}=\frac{k}{f^{\prime}\left(  R\right)  }$ is the
effective time-varying gravitational \textquotedblleft
constant\textquotedblright.

The effective energy-momentum tensor $T_{\mu\nu}^{f\left(  R\right)  }$ is
defined as
\begin{equation}
T_{\mu\nu}^{f\left(  R\right)  }=\left(  \nabla_{\mu}\nabla_{\nu}-g_{\mu\nu
}\nabla_{\sigma}\nabla^{\sigma}\right)  f^{\prime}+\frac{1}{2}\left(
f-Rf^{\prime}\right)  g_{\mu\nu}, \label{ac.04}%
\end{equation}
and it attributes the additional degrees of freedom provided by the
modification of the Einstein-Hilbert action.

In cosmological studies, we assume that on large scales, the universe is
homogeneous and isotropic described by the spatially flat FLRW line element
\begin{equation}
ds^{2}=-dt^{2}+a^{2}\left(  t\right)  \left(  dx^{2}+dy^{2}+dz^{2}\right), 
\label{fr.01}%
\end{equation}
where $a\left(  t\right)  $ is the scalar factor, and $H=\frac{\dot{a}}{a}$ is
the Hubble function and $\dot{a}=\frac{da}{dt}$.

For the line element (\ref{fr.01}) we calculate
\begin{equation}
R=6\dot{H}+12H^{2}. \label{ac.00}%
\end{equation}
Thus by substituting in (\ref{ac.02}) we end with the following system
\begin{equation}
3f^{\prime}H^{2}=k\rho_{m}+\frac{f^{\prime}R-f}{2}-3Hf^{\prime\prime}\dot{R},
\label{ac.05}%
\end{equation}%
\begin{equation}
2f^{\prime}\dot{H}+3f^{\prime}H^{2}=-2Hf^{\prime\prime}\dot{R}-\left(
f^{\prime\prime\prime}\dot{R}^{2}+f^{\prime\prime}\ddot{R}\right)
-\frac{f-Rf^{\prime}}{2}-kp_{m}. \label{ac.06}%
\end{equation}
Remark that $\rho_{m}=T_{\mu\nu}u^{\mu}u^{\nu}$ and $p_{m}=T_{\mu\nu}\left(
g^{\mu\nu}+u^{\mu}u^{\nu}\right)  $ correspond to the energy density and
pressure of the matter source, where $u^{\mu}$ is the normalised four-velocity
vector. In the following analysis, we consider the vacuum case, that is
$T_{\mu\nu}=0$. At this point, it is important to mention that all
equations are independent now. Indeed equation (\ref{ac.05}) is a constraint
equation and describes a conservation law for the dynamical system.

From the field equations (\ref{ac.05}), (\ref{ac.06}) we can express the
energy-momentum and pressure components for the effective energy-momentum
tensor $T_{\mu\nu}^{f\left(  R\right)  }$ as follow \cite{Sotiriou}
\begin{equation}
\rho_{f}=\frac{f^{\prime}R-f}{2}-3Hf^{\prime\prime}\dot{R}, \label{ac.07}%
\end{equation}%
\begin{equation}
p_{f}=2Hf^{\prime\prime}\dot{R}+\left(  f^{\prime\prime\prime}\dot{R}%
^{2}+f^{\prime\prime}\ddot{R}\right)  +\frac{f-Rf^{\prime}}{2}, \label{ac.08}%
\end{equation}
where now the equation of state (EoS) parameter, $w_{f}=\frac{p_{f}}{\rho_{f}}$,
for the effective fluid is defined as
\begin{equation}
w_{f}=-\frac{\left(  f-Rf^{\prime}\right)  +4Hf^{\prime\prime}\dot{R}+2\left(
f^{\prime\prime\prime}\dot{R}^{2}+f^{\prime\prime}\ddot{R}\right)  }{\left(
f-Rf^{\prime}\right)  +6Hf^{\prime\prime}\dot{R}}. \label{ac.09}%
\end{equation}

Many functional forms of the $f\left(  R\right)  $ have
been proposed in the literature. Some of these models are presented in Table
\ref{tab1}.%

\begin{table}[tbp] \centering
\caption{Proposed $f(R)$ models of gravity}%
\begin{tabular}
[c]{cc}\hline\hline
$\mathbf{f}\left(  R\right)  $ & \textbf{Reference}\\
$R^{\kappa}$ & \cite{dun}\\
$R+qR^{2}$ & \cite{star}\\
$R+qR^{\kappa}$ & \cite{Carrol,Noi}\\
$R+qR^{\kappa}+\beta R^{2}$ & \cite{exts}\\
$R-qR_{c}\frac{(R/R_{c})^{2n}}{(R/R_{c})^{2n}+1}$ & \cite{Hu07}\\
$(R^{b}-2\Lambda)^{c}$ & \cite{AmeTsuj}\\\hline\hline
\end{tabular}
\label{tab1}%
\end{table}%

\section{Singularity analysis}

\label{sec3}

The modern treatment of the singularity analysis is summarised in an algorithm
proposed by Ablowitz, Ramani and Segur, known as the ARS algorithm
\cite{Abl1, Abl2, Abl3}. In the following lines, we present the three basic steps
of the ARS algorithm for ordinary differential equations. First, we refer the reader to \cite{buntis} for a pedagogical review.

We assume now an ordinary differential equation $\mathcal{F}\left(
t,x\left(  t\right), \dot{x}\left(  t\right), \ddot{x}\left(  t\right)
,...\right)  \equiv0$ where $t$ is the independent variable and $x\left(
t\right)  $ is the dependent variable. The first step of the ARS algorithm is
to determine whether a moveable singularity exists. For that, we assume that
near the singularity, the solution of the differential equation is
asymptotically described by the function
\begin{equation}
x\left(  t\right)  =x_{0}\left(  t-t_{0}\right)  ^{p}, \label{sd1}%
\end{equation}
where $t_{0}$ marks the position of the singularity. Hence by substituting
this functional form into the differential equation from the dominant power
terms we derive the two quantities, $p$ and $x_{0}$. For the exponent $p$, 
it should be a negative integer number for the singularity to be a pole. However, this has since been relaxed to include fractional exponents, even positive ones, 
as the derivative of a positive fractional exponent eventually gives a negative 
exponent and then gives a singularity.

The second step of the ARS algorithm is related to studying the
existence of conservation laws, that is, of the integration constants. A
$d$th-order ordinary differential equation requires $d$ integration constants. In contrast, one of the integration constants is the position of the movable singularity,
that is, the parameter $t_{0}$, we should prove the existence of $d-1$ integration
constants.\ To do so, we substitute
\begin{equation}
x\left(  t\right)  =x_0(t-t_{0})^{p}+m(t-t_{0})^{p+s} \label{sd2}%
\end{equation}
into the dominant terms of the equation and collect the terms linear in $m$ where the coefficient first enters the expansion. When the multiplier of $m$ is zero, the value of $m$ is arbitrary. The latter requirement provides a
polynomial equation in terms of $s$. \ The zeros of the polynomial give the
values of the resonances $s$. One of these must be $-1$, which is associated
with the integration constant $t_{0}$, while the remaining $s$ should be rational numbers.

The third step of the algorithm is known as the consistency test. Specifically,
we substitute an appropriate Laurent series into the ordinary differential equation to ensure that there is consistency with the type of
Series, which the analysis of the dominant terms implies. That means that we
validate the existence of the solution.

When a given differential equation passes the ARS algorithm, we say it
possesses the Painlev\'{e} property and is integrable. The
analytic solution is expressed in terms of a Laurent expansion.

In order to demonstrate the application of the ARS algorithm to ordinary
differential equations we consider the Painlev\'{e}-Ince Equation
\cite{Mahomed85}
\begin{equation}
\ddot{x}+3x\dot{x}+x^{3}=0.
\end{equation}

For the first step of the ARS algorithm, we substitute expression (\ref{sd1})
into the differential equation to obtain
\[
x_{0}p\left(  p-1\right)  \left(  t-t_{0}\right)  ^{p-2}+3p x_{0}^2\left(
t-t_{0}\right)  ^{2p-1}+x_{0}^{3}\left(  t-t_{0}\right)  ^{3p}=0,
\]
from which it follows that the singular behaviour is for $p=-1$, and $(x_{0}-2) (x_{0}-1) x_{0}=0$, whence, $x_{0}=1$
or $x_{0}=2$.

We proceed with the second step of the algorithm. We replace expression
(\ref{sd2}) into the Painlev\'{e}-Ince Equation and we linearise around $m$.  Thus
the coefficient of $m$ provides the polynomial
\begin{equation}
P\left(  s;x_{0}\right)  =s^{2}+3s\left(  x_{0}-1\right)  +3x_{0}\left(
x_{0}-2\right)  +2,
\end{equation}
where $P\left(  s;1\right)  =0$ gives $s=-1,~s=1$ and $P\left(  s,2\right)=0  $
gives $s=-1$ and $s=-2$. The latter means that there are two Laurent
expansions that may solve the Painlev\'{e}-Ince Equation.

Assume now the case for $x_{0}=1$; then we consider the Laurent expansion \[
x\left(  t\right)  =\left(  t-t_{0}\right)  ^{-1}+x_{1}\left(  t-t_{0}\right)
^{0}+x_{2}\left(  t-t_{0}\right)  ^{1}+x_{3}\left(  t-t_{0}\right)  ^{2}...~,
\]
we replace the original equation, and we find that it is a solution for
$x_{1}$ arbitrary, which is the second integration constant of the problem,
and $x_{2}=-x_{1}^{2}$ $,~x_{3}=x_{1}^{3}$, etc.

We now proceed with the application of the singularity analysis in
$f\left(  R\right)  $-cosmology.

\section{Analytic solutions}

\label{sec4}

In order to apply the singularity analysis, we substitute (\ref{ac.00}) into
(\ref{ac.05}).  Thus we obtain a second-order equation for the Hubble function
$H\left(  t\right)  $. In the following, we consider some power-law $f\left(
R\right)  $ models.

\subsection{Model $f\left(  R\right)  =R^{\kappa}$}

As a toy model we assume the power-law model $f\left(  R\right)  =R^{\kappa
},~\kappa\neq0,1$. Thus, equation (\ref{ac.05}) is%
\begin{equation}
\kappa\left(  \kappa-1\right)  H\ddot{H}+\left(  \kappa-1\right)  \dot{H}%
^{2}+\left(  \kappa\left(  7-4\kappa\right)  -4\right)  H^{2}\dot{H}+\left(
2\left(  \kappa-2\right)  \right)  H^{4}=0. \label{qe.01}%
\end{equation}

We follow the ARS algorithm as described in the previous section. We find that
the leading-order behaviour is
\begin{equation}
H\left(  t\right)  =\frac{H_{0}}{t}~,~H_{0}=\frac{1}{2}~ \text{or}~H_{0}=\frac{\left(
\kappa-1\right)  \left(  2\kappa-1\right)  }{2-\kappa},
\end{equation}
where without loss of generality, we assume that the singularity is at
$t_{0}=0$.

Hence, the leading-order behaviour corresponds to the power-law scale factor
$a_{0}=t^{H_{0}}$, which describes the ideal gas solution with an equation of state
parameter $w_{eff}\left(  H_{0}\right)  =\frac{2}{3H_{0}}-1$. Specifically,
$w_{eff}\left(  \frac{1}{2}\right)  =\frac{1}{3}$, which corresponds to the
radiation asymptotic solution as described by dynamical analysis
\cite{fran1}. Remark that for $\kappa=\frac{1}{2}$ there exists only one
singular behaviour, that with $H_{0}=\frac{1}{2}$.

We continue with the second step of the ARS algorithm; the resonances are
\begin{equation}
H_{0}=\frac{1}{2}:~s_{1}^{A}=-1,~s_{2}^{A}=\frac{5-4\kappa}{2\left(
1-\kappa\right)  }~, \label{ee.01}%
\end{equation}%
\begin{equation}
H_{0}=\frac{\left(  \kappa-1\right)  \left(  2\kappa-1\right)  }{2-\kappa
}:~s_{1}^{B}=-1,~s_{2}^{B}=\frac{\left(  2\kappa-1\right)  \left(
4\kappa-5\right)  }{\kappa-2}~. \label{ee.02}%
\end{equation}

Consequently, when $\kappa$ is a rational number, the resonances are rational
numbers. From the sign of the second resonance, we can determine the direction
of the Laurent expansion, which describe the analytic solution. Indeed,
for $s_{2}\geq0$, the analytic solution is described by a Right Laurent
expansion, while for $s_{2}<0$, the analytic solution is described by a Left
Laurent expansion. In addition, $s_{2}^{A}>0$ for $\kappa<1$ and $\kappa
>\frac{5}{4}$, while $s_{2}^{B}>0$ for $\frac{1}{2}<\kappa<\frac{5}{4}$ and
$\kappa>2$. The value $s_{2}=0$ means that the second integration constant is
the coefficient term of the leading-order behaviour. However, from our analysis, it
is clear that $H_{0}$ cannot be arbitrary. Thus, the singularity analysis
fails for $\kappa=\frac{5}{4}$.

We proceed with the final step of the ARS algorithm, the consistency test, by selecting specific values for the free parameter $\kappa$.

Assume now that $\kappa=\frac{1}{2}$, then, for the leading-order behaviour
with $H_{0}=\frac{1}{2}$, we find the resonances $s_{1}^{A}=-1,~s_{2}^{A}=3$.
We write the Laurent expansion%
\begin{equation}
H\left(  t\right)  =\frac{1}{2}t^{-1}+H_{1}+H_{2}t+H_{3}t^{2}+H_{4}t^{3}+...~
\label{qe.02}%
\end{equation}
and, by replacing in (\ref{qe.01}), it follows that $H_{1}=0,~H_{2}=0,~H_{4}%
=0$,~$H_{5}=0$, $H_{6}=\frac{22}{7}H_{3}^{2}$, in which $H_{3}$ is the second
integration constant.

Assume the cosmological model with $\kappa=\frac{3}{2}$. It follows that
$H_{0}=\frac{1}{2}:~s_{1}^{A}=-1,~s_{2}^{A}=1$ and $H_{0}=2:~s_{1}%
^{B}=-1,~s_{2}^{B}=-4$. For the case with $H_{0}=\frac{1}{2}$, we substitute
the Right Laurent expansion (\ref{qe.02}) into equation (\ref{qe.01}) and it
follows that $H_{1}$ is an arbitrary constant and $H_{2}=-\frac{2}{9}H_{1}%
^{2},$ etc.

For the case with $H_{0}=2$, we write the Left expansion%
\begin{equation}
H\left(  t\right)  =...+H_{-3}t^{-4}+H_{-2}t^{-3}+H_{-1}t^{-2}+\frac{1}%
{2}t^{-1}, \label{qe.02a}%
\end{equation}
where it follows that~$H_{-1}\,\ $\ is arbitrary and $H_{-2}=\frac{1}{2}\left(
H_{-1}\right)  ^{2},~H_{-3}=\frac{1}{4}\left(  H_{-1}\right)  ^{3}$, etc.

We summarise the results in the following Proposition.

\textbf{Proposition 1: }\textit{For the }$f\left(  R\right)  =R^{\kappa
},~\kappa\neq0,1,~$\textit{the cosmological field equations possess the
Painlev\'{e} property for }$\kappa\neq\frac{5}{4}$\textit{, thus the field
equations are integrable.}

Of course, the model $f\left(  R\right)  =R^{\kappa}$ is a straightforward cosmological scenario where according to the analysis presented in
\cite{fran1}, it is not cosmologically viable. In addition, from the previous analysis, we know that this power-law model gives an integrable set of cosmological field equations with conservation laws provided by the symmetry analysis.

\subsection{Model $f\left(  R\right)  =R+qR^{\kappa}-2\Lambda$}

We now proceed with the second model of our analysis, the $f\left(
R\right)  =R+qR^{\kappa}-2\Lambda$,~$\kappa\neq0,1$. For $\Lambda=0$, the
singularity analysis for this model has been applied in \cite{anst2}. It
was found that the cosmological field equations possess the Painlev\'e property
for $\kappa>2$. The analysis has been performed for the system of second-order
equations (\ref{ac.00}) and (\ref{ac.06}) for the variables $\left(
a, R\right)  $. \ In this attempt, we proceed with the analysis for the
second-order ordinary differential equation (\ref{ac.05}) for the variable
$H$, where we asssume $\Lambda\neq0$.

Thus, for this specific $f\left(  R\right)  $ model, the master equation for
the Hubble function reads%
\begin{align}
0  & =\frac{\Lambda}{3}-H^{2}+6^{\kappa-1}q\left(  2H^{2}+\dot{H}\right)
^{\kappa-2}\left(  2\left(  \kappa-2\right)  H^{4}+\left(  \kappa-1\right)
\dot{H}^{2}-\left(  \kappa-1\right)  \kappa H\ddot{H}\right)  \label{qe.03}\\
& -6^{\kappa-1}q\left(  2H^{2}+\dot{H}\right)  ^{\kappa-2}\left(  6+\left(
9+\left(  \kappa-2\right)  \right)  \left(  \kappa-2\right)  \right)  \dot
{H}H^{2}.\nonumber
\end{align}

Hence, by replacing $H\left(  t\right)  =H_{0}t^{p}$, we find that the dominant
terms are these provided by the $R^{\kappa}$ component for $\kappa>2$, and
$p=-1$, $H_{0}=\frac{1}{2}$ or $H_{0}=\frac{\left(  \kappa-1\right)  \left(
2\kappa-1\right)  }{2-\kappa}$. From the second step of the ARS\ algorithm, we
find that the resonances are those given in expressions (\ref{ee.01}) and
(\ref{ee.02}) respectively.

For the consistency test, we assume the case $\kappa=3$, for which the resonances
are $s_{1}=-1$ and $s_{2}=35$. We write the Laurent expansion
\begin{equation}
H\left(  t\right)  =-10t^{-1}+H_{1}t^{0}+H_{2}t^{1}+H_{3}t^{2}+...~~,~
\label{qe.04}%
\end{equation}
which corresponds to the leading-order behaviour for the solution
with$~H_{0}=\frac{\left(  \kappa-1\right)  \left(  2\kappa-1\right)
}{2-\kappa}$. Then by replacing in (\ref{qe.03}) we find that satisfies the
consistency test with nonzero coefficients the $H_{4}$, $H_{6}$, $H_{8}$ etc
and second integration constant the $H_{35}$.

Similarly, for the leading-order behaviour with $H_{0}=\frac{1}{2}$, and
$\kappa=3$, we write the Laurent expansion%
\begin{equation}
H\left(  t\right)  =\frac{1}{2}t^{-1}+H_{1}t^{-1+\frac{1}{4}}+H_{2}t^{-1+\frac{2}{4}}+H_{3}t^{-1+\frac{3}{4}}+...~,
\label{qe.05}%
\end{equation}
where we find that from $H_{1}$ to $H_{10}$,  where $H_{7}$ is arbitrary, that is, it is
the second integration constant and the only nonzero coefficient is
$H_{9}=-\frac{2}{3861 q H_{7}}$.

\textbf{Proposition 2: }\textit{For the }$f\left(  R\right)  =R+qR^{\kappa
}-2\Lambda,~\kappa\neq0,1,~$\textit{the cosmological field equations possess
the Painlev\'{e} property for }$\kappa>2$.\textit{  Thus, the field equations
are integrable.}

\subsection{Model $f\left(  R\right)  =R+qR^{\kappa}+\beta R^{2}-2\Lambda$}

We now assume the power-law extension for the quadratic inflationary model
with $f\left(  R\right)  =R+qR^{\kappa}+\beta R^{2}-2\Lambda$. We apply the
ARS algorithm as before, and we find that the field equations possess the
Painlev\'{e} property for $\kappa>2$, with leading-order behaviour and resonances
as given by (\ref{ee.01}) and (\ref{ee.02}).  That means that the singularity
is described by the dominant terms given by the $R^{\kappa}$.

We now present the case $\kappa=3$ and substitute the two Laurent
expansions (\ref{qe.04}), (\ref{qe.05}) into the master equation for the Hubble function.

It follows that the Laurent expansion (\ref{qe.04}) solves the master equation
(\ref{ac.05}) with arbitrary coefficient being $H_{35}$, and
\begin{equation}
H_{1}=0,~H_{2}=-\frac{\beta}{1188q},~H_{3}=0,~H_{4}=\frac{437\beta^{2}%
-4356q}{3062616480q^{2}},~etc~.
\end{equation}

In \ a similar way, expansion (\ref{qe.05}) is a solution for the master
equation with nonzero coefficients being $H_{7}$ which is an integration
constant, $H_{8}=-\frac{\beta}{54q},~H_{9}=-\frac{2\left(  3q-\beta
^{2}\right)  }{11583 q^{2} H_{7}}$, etc. Thus the following Proposition follows.

\textbf{Proposition 3: }\textit{The cosmological field equations in
}$\mathit{f}\left(  R\right)  =R+qR^{\kappa}+\beta R^{2}-2\Lambda,~\kappa
\neq0,1,$ theory \textit{ possess the Painlev\'{e} property when }$\kappa
>  2$\textit{, and Laurent expansions express the analytic solutions.}

\subsection{Model $f\left(  R\right)  =\left(  R-2\Lambda\right)  ^{\kappa}$}

For the cosmological model $f\left(  R\right)  =\left(  R-2\Lambda\right)
^{\kappa}$, the master equation for the Hubble function is%
\begin{align}
0  &  =9\kappa\left(  \kappa-1\right)  H\ddot{H}-12\left(  \kappa-2\right)
H^{4}+\left(  \Lambda-3\dot{H}\right)  \left(  \Lambda+3\left(  \kappa
-1\right)  \dot{H}\right)  +\nonumber\\
&  +3H^{2}\left(  \Lambda\left(  \kappa-4\right)  +3\left(  4+\kappa\left(
4\kappa-7\right)  \right)  \dot{H}\right). \label{qe.06}%
\end{align}

The application of the two first steps for the ARS algorithm gives that the
leading-order terms and the resonances are those given by expressions
(\ref{qe.04}), (\ref{qe.05}). We assume specific values for the parameter
$\kappa$ in order to perform the consistency test.

For $\kappa=\frac{1}{2}$ and the Laurent expansion (\ref{qe.02}) from equation
(\ref{qe.06}) it follows that $H_{3}$ is arbitrary and $H_{1}=0,$~$H_{2}%
=\frac{\Lambda}{9}$, $H_{4}=-\frac{2}{405}\Lambda^{2}$, $H_{5}=-\frac{4}%
{9}\Lambda H_{3}$, etc.

Similarly, for $\kappa=\frac{3}{2}$ and the Laurent expansion (\ref{qe.02a}) we
find that it does not solve the master equation (\ref{qe.06}), which means that
the ARS algorithm fails for this branch.

We conclude that the field equations in this theory are integrable with
dominant term and resonances as given by expression (\ref{qe.04}).

\textbf{Proposition 4: }\textit{The cosmological field equations in the
}$f\left(  R\right)  =\left(  R-2\Lambda\right)  ^{\kappa},~\kappa\neq
0,1,$\textit{ possess the Painlev\'{e} property, and Laurent expansions express the analytic solutions.}

From the above results, we can make the conjecture that 

\textbf{Conjecture 5: }\textit{For every $f\left(
R\right)  $ model, near to the movable singular behaviour, the
$f\left(  R\right)  $ function is dominated by the $R^{\kappa}$ term, the
field equations possess the Painlev\'{e} property for specific values of
$\kappa$, with leading-order behaviour corresponding to the radiation solution and the ideal
gas solution.}

The latter result is important for better understanding by using
analytic techniques, the behaviour of a given cosmological model around a
specific cosmological epoch.

\section{Conclusions}

\label{sec5}

The singularity analysis systematically investigates the
integrability properties of dynamical systems and constructs analytic
solutions. The singularity analysis should be seen as a complementary rather
than competitive symmetry analysis method for studying dynamical
systems. These two approaches are powerful tools for the study of
integrability. Regarding the modified theories of gravity, symmetry
analysis has been used to constrain the theory's unknown functions and
parameters. For instance, in $f\left(  R\right)  $-gravity, the
symmetry analysis provides constraints for the functional form of the
$f\left(  R\right)  $ function. However, this substantial property is partially
true for the singularity analysis. As we show from our analysis, only the free
parameters of a specific $f\left(  R\right)  $ function were constrained by the
requirement of the ARS algorithm.

In this study, we applied the singularity analysis in four power-law $f\left(
R\right)  $ models which have been proposed before in the literature. We
considered the power-law $f\left(  R\right)  =R^{\kappa}$ theory as a toy
study model. It is well-known that $f\left(  R\right)  =R^{\kappa}$
cosmology is an integrable model. However, we proved that the field equations
possess the Painlev\'{e} property. Specifically, there are two possible
Laurent expansions describe the analytic solutions where the dominant
terms near the singularity describe a radiation-dominated universe or an
ideal gas-dominated universe with an equation of state parameters depending upon the
parameter $\kappa$. The results for the  $f\left(  R\right)  =R^{\kappa}$ model were used to analyse the power-law models.

We considered the following models, $f\left(  R\right)  =R+qR^{\kappa
}-2\Lambda$, $f\left(  R\right)  =R+qR^{\kappa}+\beta R^{2}-2\Lambda$ and
$f\left(  R\right)  =\left(  R-2\Lambda\right)  ^{\kappa}$, for which we found
that the resulting master equations for the Hubble function possess the
Painlev\'{e} property for specific values of $\kappa$, with dominant terms
that of $f\left(  R\right)  =R^{\kappa}$ and the analytic solutions can be
expressed by left or right Laurent expansions. Finally, we calculated the
analytic solution for the field equations for specific values of $\kappa.$

From the above results, we conjectured that,  near to the movable singular behaviour, any
$f\left(  R\right)  $ function is dominated by the $R^{\kappa}$ term, whence the
field equations possess the Painlev\'{e} property for specific values of
$\kappa$, with leading-order behaviour corresponding to the radiation solution and the ideal
gas solution. 

The latter result is important for better understanding by using
analytic techniques, the behaviour of a given cosmological model around a
specific cosmological epoch. At this point, it is important to recall that the
asymptotic solution $H\left(  t\right)  =\frac{H_{0}}{t}$ corresponds
to the ideal gas exact solution. For $H_{0}=\frac{1}{2}$ the asymptotic
solution describes the radiation era, while for $H_{0}=\frac{2}{3}$
the matter-dominated era is recovered. For the power-law model $f\left(
R\right)  =R^{\kappa}$, for $\kappa>2$, we found two
leading-order behaviours, which describe the radiation era and the ideal gas
solution with equation of state parameter $w_{eff}\left(  \kappa\right)
=\frac{7\kappa+1-6\kappa^{2}}{\left(  \kappa-1\right)  \left(  2\kappa
-1\right)  }$, from where it follows that for $\kappa>2$,
$w_{eff}\left(  \kappa\right)  <-1$ and $w_{eff}\left(
\kappa\rightarrow+\infty\right)  \simeq-1$. Consequently, the second analytic solution can describe a solution with initial conditions near to the
inflationary epoch, or to the late-time acceleration phase of the universe.
The latter comments hold also for the $f\left(  R\right)  =R+qR^{\kappa
}-2\Lambda$ model as also for every $f\left(  R\right)  $
function dominated by the power-law $R^{\kappa}$ term near the
singularity, as described by Conjecture 5. An application of the above
solutions are to use them as exact background spacetimes for studying the
cosmological perturbations around specific cosmological epochs, such as the
inflationary era. 

In future work, we plan to extend this specific analysis for the study of
other higher-order gravitational theories.

\section*{Data Availability Statements}

Data sharing not applicable to this article as no datasets were generated or analysed during the current study.

\section*{Author contributions}

\textbf{Genly Leon}: Writing—review and editing; investigation. \textbf{Andronikos Paliathanasis}: Methodology; formal analysis;
writing—original draft; investigation. \textbf{Peter G. L. Leach}: Supervision; writing—review and editing; investigation.

\begin{acknowledgments}
Genly Leon was funded by Vicerrectoría de Investigación y Desarrollo Tecnológico (Vridt) at Universidad Católica del Norte (UCN) through
Concurso De Pasantías De Investigación Año 2022, Resolución Vridt N° 040/2022 and through Resolución Vridt N° 054/2022. He also thanks the support of Núcleo de Investigación Geometría Diferencial y Aplicaciones, Resolución Vridt N°096/2022.

\end{acknowledgments}

\section*{Conflict of interest statement}

This work does not have any conflicts of interest.


\begin{thebibliography}{99}                                                                                               %


\bibitem {ar1}F.M. Arscott, Periodic Differential Equations, Pergamon Press,
Oxford, (1964)

\bibitem {v2}E. Kasner, Am. J. Math. 43, 217 (1921)

\bibitem {v3}T. Christodoulakis and P.A.\ Terzis, J. Phys.: Conf. Ser. 68,
012039 (2007)

\bibitem {v4}T. Christodoulakis, G. Kofinas, E. Korfiatis, G.O. Papadopoulos
and A. Paschos, J.Math.Phys. 42, 3580 (2001)

\bibitem {v5}P.A. Terzis and T. Christodoulakis, Gen.\ Relat. Gravit. 41, 469 (2009)

\bibitem {nl1}A. Paliathanasis,\ S. Capozziello and P.G.L. Leach, Phys. Lett.
B 755, 8 (2016)

\bibitem {sm1}M.C. Kweyama, K.S. Govinder and S.D. Maharaj, Class. Quantum
Grav. 28, 105005 (2011)

\bibitem {sm2}T. Christodoulakis, N. Dimakis and P.A. Terzis, J. Phys. A:
Math.\ Theor. 47, 095202 (2014)

\bibitem {sm3}S. Cotsakis, P.G.L.\ Leach and H. Pantazi, Grav. Cosmol. 4, 314 (1998)

\bibitem {ap1}T. Pailas, N. Dimakis, A. Paliathanasis, P.A.\ Terzis and
T.\ Christodoulakis, Phys. Rev. D 102, 063524 (2020)

\bibitem {sym4}S. Dussault and V. Faraoni, EPJC 80, 1002 (2020)

\bibitem {ns1}R. de Ritis, G. Marmo, G. Platania, C. Rubano, P. Scudellaro and
C. Stornaiolo, Phys. Rev. D. 42, 1091 (1990)

\bibitem {ns2}A.\ Paliathanasis, M.\ Tsamparlis and S. Basilakos, Phys. Rev. D
90 103524 (2014)

\bibitem {ns4}Y. Zhang, Y.-G. Gong, Z.-H. Zhu, Phys. Lett. B 688, 13 (2010)

\bibitem {ns5}B. Vakili and F. Khazaie, Class.\ Quantum Gravit. 29, 035015 (2012)

\bibitem {ns6}K. Atazadeh and F.\ Darabi, EPJC 72, 2016 (2012)

\bibitem {pa1}P. Painlev\'{e}, Le\c{c}ons sur la th\'{e}orie analytique des
\'{e}quations diff\'{e}rentielles (Le\c{c}ons de Stockholm, 1895) (Hermann,
Paris, 1897). Reprinted, O$\!$euvres de Paul Painlev\'{e}, vol.~I,
\'{E}ditions du CNRS, Paris, (1973).

\bibitem {pa2}P. Painlev\'{e}, M\'{e}moire sur les \'{e}quations
diff\'{e}rentielles du second ordre dont l'int\'{e}grale g\'{e}n\'{e}rale est
uniforme Bulletin of the Mathematical Society of France 28, 201 (1900)

\bibitem {pa3}P. Painlev\'{e}, Sur les \'{e}quations diff\'{e}rentielles du
second ordre et d'ordre sup\'{e}rieur dont l'int\'{e}grale g\'{e}n\'{e}rale
est uniforme Acta Mathematica 25, 1 (1902)

\bibitem {pa4}P. Painlev\'{e}, Sur les \'{e}quations diff\'{e}rentielles du
second ordre \`{a} points critiques fixes Comptes Rendus de la Acad\'{e}mie
des Sciences de Paris 143, 1111 (1906)

\bibitem {Kowlevskaya89a}S. Kowalevski, Sur la probl\`{e}me de la rotation
d'un corps solide autour d'un point fixe, Acta. Math. 12, 177 (1889)

\bibitem {cont1}R. Conte, The Painlev\'e Property: One Century Later, CRM Series
in Mathematical Physics, Springer-Verlag, New York (1999)

\bibitem {CotsakisLeach}S. Cotsakis and P.G.L. Leach, J. Phys. A: Math. Gen.
27, 1625 (1994)

\bibitem {Con1}G. Contopoulos, B. Grammaticos and A. Ramani, J. Math. Phys A
Math. Gen. 25, 5795 (1993)

\bibitem {Demaret}J. Demaret and C. Scheen, J. Math. Phys. A: Math. Gen. 29,
59 (1996)

\bibitem {bun1}F. Christiansen, H.H. Rugh and S.E. Rugh, J. Phys. A: Math.
Gen. 28, 657 (1995)

\bibitem {miritzis}J. Miritzis, P.G.L. Leach and S. Cotsakis, Grav. Cosmol. 6,
282 (2000)

\bibitem {ftAn}A. Paliathanasis, J.D.\ Barrow and P.G.L. Leach, Phys. Rev. D
94, 023525 (2016)

\bibitem {fqAn}W. Khyllep, A. Paliathanasis and J. Dutta, Phys. Rev. D\ 103,
103521 (2021)

\bibitem {cots11}S. Cotsakis, G. Kolionis and A.\ Tsokaros, Phys. Lett. B 721,
1 (2013)

\bibitem {cots2}S. Cotsakis, S. Kadry, G. Kolionis and A. Tsokaros, Phys.
Lett. B 755, 387 (2016)

\bibitem {cots}S. Cotsakis, J. Demaret, Y. De Rop and L. Querella, Phys. Rev.
D 48, 4595 (1993)

\bibitem {sbs}S. Basilakos, A. Paliathanasis, J.D. Barrow and
G.\ Papagiannopoulos, EPJC 78, 684 (2018)

\bibitem {Buda}H.A. Buchdahl, Mon. Not. Roy. Astron. Soc. 150 1 (1970)

\bibitem {capQ}S.\ Capozziello, Int. J Mod. Phys. D 11 483 (2002)

\bibitem {Sotiriou}T.P. Sotiriou and V. Faraoni Rev. Mod. Phys. 82 451 (2010)

\bibitem {odin1}S. Nojiri and S.D. Odintsov, Phys. Rep. 505 59 (2011)

\bibitem {qua1}H. Nariai and K. Tomita, Prog. Theor. Phys. 46, 776 (1971)

\bibitem {qua2}G.V. Bicknell, J. Phys. A.: Math. Nucl. Gen. 7, 1061 (1974)

\bibitem {qua3}J.D. Barrow, Nucl. Phys. B 296, 679 (1988)

\bibitem {star}A.A. Starobinsky, Phys. Lett. B 91, 99 (1980)

\bibitem {planck2015}P.A.R. Ade et al. (Planck 2015 Collaboration), A.\&A.
594, A20 (2016)

\bibitem {anst1}A. Paliathanasis, EPJC 77, 027 (2017)

\bibitem {anst2}A. Paliathanasis and P.G.L.\ Leach, Phys. Lett. A\ 380, 2815 (2016)

\bibitem {dun}N. Goheer, J. Larena and P.K.S. Dunsby, Phys. Rev. D 80, 061301 (2009)

\bibitem {Carrol}S. M. Carrol, V. Duvvuri, M. Trodden and M. S. Turner, Phys.
Rev. D., 70, 043528 (2004);

\bibitem {Noi}S. Nojiri and S, D. Odintsov, Phys. Rev. D., 73, 124038 (2006)

\bibitem {exts}D.Y. Cheng, H.M. Lee and S.C. Park, Phys. Lett. B 805, 135456 (2020)

\bibitem {Hu07}W. Hu and I. Sawicki, Phys. Rev. D., 76, 064004 (2007)

\bibitem {AmeTsuj}L. Amendola and S. Tsujikawa, Phys. Lett. B., 660, 125 (2008)

\bibitem {Abl1}M.J. Ablowitz, A. Ramani and H. Segur, \ Lettere al Nuovo
Cimento 23, 333 (1978)

\bibitem {Abl2}M.J. Ablowitz, A. Ramani and H. Segur, J. Math. Phys. 21, 715 (1980)

\bibitem {Abl3}M.J. Ablowitz, A. Ramani and H. Segur, J. Math. Phys. 21, 1006 (1980)

\bibitem {buntis}A. Ramani, B. Grammaticos and T. Bountis, Physics Reports,
180 159 (1989)

\bibitem {Mahomed85}F.M. Mahomed and P.G.L. Leach, Qu\ae stiones
Mathematic\ae \ 8, 241 (1985)


\bibitem {fran1}L. Amendola, D. Polarski and S. Tsujikawa, Int. J. Mod. Phys.
D 16, 1555 (2007)
\end{thebibliography}
\end{document}